\documentclass[traditabstract]{aa} 
\usepackage{longtable}
\usepackage{supertabular}
\usepackage{graphicx}
\usepackage{array}
\usepackage{tabularx}
\usepackage{ragged2e}
\usepackage{color}
\usepackage{natbib}
\usepackage{multirow}
\usepackage{multicol}
\usepackage{wasysym}
\usepackage[varg]{txfonts}
\usepackage{lscape}

\begin{document}

   \title{Forbidden oxygen lines at various nucleocentric distances in comets\thanks{Based on observations made with ESO Telescope at the La Silla Paranal Observatory under programs ID 073.C-0525, 277.C-5016, 080.C-0615 and 086.C-0958.}}

   \author{A. Decock
          \inst{1}
			,
          E. Jehin\inst{1}	
          		,
		P. Rousselot\inst{2}
		,
			D. Hutsem\'ekers\inst{1}
			,
			J. Manfroid\inst{1}
			,
			S. Raghuram\inst{3,4}
			,
			A. Bhardwaj\inst{3}
			,
			B. Hubert\inst{1}
          }

   \institute{Institut d'Astrophysique, de G\'eophysique et Oc\'eanographie, Universit\'e de Li\`ege, All\'ee du 6 ao\^ut 17, 4000 Li\`ege, Belgium.\\
   	              \email{adecock@ulg.ac.be}
	              \and
	              University of Franche-Comt\'e, Observatoire des Sciences de l'Univers THETA, Institut UTINAM - UMR CNRS 6213, BP 1615, 25010. Besan\c con Cedex, France.
	              \and
	              Space Physics Laboratory, Vikram Sarabhai Space Centre, Trivandrum 695 022, India.
	              \and
	              Department of Physics, Imperial College London, London SW7 2AZ, UK.
             }

   \date{Received ; accepted }

  \abstract
   {}
   {To study the formation of the [OI] lines -~i.e., 5577.339~$\AA$ (the green line), 6300.304~$\AA$ and 6363.776~$\AA$ (the two red lines)~- in the coma of comets and to determine the parent species of the oxygen atoms using the green to red-doublet emission intensity ratio, I$_{5577}$/(I$_{6300}$+I$_{6364}$), (hereafter the G/R ratio) and the lines velocity widths.}
   {We acquired at the ESO Very Large Telescope high-resolution spectroscopic observations of comets C/2002 T7 (LINEAR), 73P-C/Schwassmann-Wachmann 3, 8P/Tuttle, and, 103P/Hartley 2 when they were close to the Earth (< 0.6 au). Using the observed spectra, which have a high spatial resolution (<60~km/pixel), we determined the intensities and the widths of the three [OI] lines. We have spatially extracted the spectra  in order to achieve the best possible resolution of about 1-2$^{\prime\prime}$, i.e., nucleocentric projected distances of 100 to 400~km depending on the geocentric distance of the comet. We have decontaminated the [OI] green line from C$_{2}$ lines blends that we have identified.}
   { It is found that the observed G/R ratio on all four comets varies as a function of nucleocentric projected distance (between $\sim$0.25 to $\sim$0.05 within 1000~km).
   This is mainly due to the collisional quenching of O($^{1}$S) and O($^{1}$D) by water molecules in the inner coma.
    The observed green emission line width is about 2.5~km~s$^{-1}$ and decreases as the distance from the nucleus increases which can be explained by the varying contribution of CO$_{2}$ to the O($^{1}$S) production in the innermost coma.
    The photodissociation of CO$_{2}$ molecules seems to produce O($^{1}$S) closer to the nucleus while the water molecule forms all the O($^{1}$S) and O($^{1}$D) atoms beyond $10^{3}$~km. Thus we conclude that the main parent species producing O($^{1}$S) and O($^{1}$D) in the inner coma is not always the same. The observations have been interpreted in the framework of the coupled-chemistry-emission model of \cite{Bhardwaj2012} and the upper limits of CO$_{2}$ relative abundances are derived from the observed G/R ratios. Measuring the [OI] lines could indeed provide a new way to determine the CO$_{2}$ relative abundance in comets. }
   {}

               
 \authorrunning{Decock et al.}
\titlerunning{[OI] lines in comets at various nucleocentric distances}

   \maketitle

\section{Introduction}

Comets are among the best preserved specimens of the primitive solar nebula as the nucleus composition did not evolve much since their formation, 4.6 billion years ago. Studying the composition of comet nuclei is thus essential to understand the formation and the evolution of material within our solar system. Oxygen is one of the most abundant elements in comets since most of the cometary ices are constituted of H$_{2}$O, CO$_{2}$ and CO molecules. All these molecules coming from the sublimation of cometary ices can produce oxygen atoms in the coma by photodissociation. Oxygen atoms in comets are analyzed through the three forbidden oxygen lines located at 5577.339~$\AA$ for the green line and at 6300.304~$\AA$ and 6363.776~$\AA$ for the red-doublet. These emission lines are due to the electronic transition of the oxygen atoms from the $^{1}$S state to the $^{1}$D state (green line) and from the $^{1}$D to the ground state (red doublet). In previous papers \citep[and references therein]{Cochran1984, Cochran2001, Cochran2008, Capria2010}, oxygen atoms are found to come mainly from the photodissociation of water molecules at a heliocentric distance around 1~au. \cite{Decock2013} analysed about 50 spectra belonging to 12 comets observed at $r$ $\sim$1~au and compared the theoretical G/R ratio values (see $^{1}$S/$^{1}$D in Table~\ref{productionrates}) with the observed ones. This work confirms that H$_{2}$O is the main parent source of atomic oxygen in the cometary coma within radial distance of 8000~km from the nucleus. The observation of these emission lines at larger heliocentric distances (>2.5~au) have shown higher G/R ratio due to significant CO$_{2}$ contribution in the production of oxygen atoms \citep{Decock2013}. A similar higher G/R ratio values on comets has also been observed by \cite{McKay2012a, McKay2013}. In the present paper, we study the three forbidden oxygen lines in the inner coma as close as possible to the nucleus in order to determine the parent species producing oxygen atoms at different nucleocentric distances. 
This analysis was made for four comets of different types (New, Jupiter Family, Halley type) observed with the ESO Very Large Telescope with a good spatial resolution and high signal-to-noise. We also study the effect of the collisional quenching of O($^{1}$S) and O($^{1}$D) with the ambient cometary species (mainly water) on the G/R ratio profile close to the nucleus ($\lesssim$~1000~km).

\begin{table}
\centering
\caption{The calculated O($^{1}$S) and O($^{1}$D) photorates and $^{1}$S/$^{1}$D ratios for the major oxygen bearing species by \cite{Raghuram2013} for quiet solar condition and at $r$~$\approx$~1~au.}
\label{productionrates}
\begin{tabular}{l || c c || c r}
\hline
\hline
Parents & \multicolumn{2}{|c||}{Emission rate (s$^{-1}$)} & Ratio  \\
 & O($^{1}$S) & O($^{1}$D) & $^{1}$S/$^{1}$D \\
\hline
H$_{2}$O & 3.78~10$^{-8}$ & 9.5~10$^{-7}$ & 0.040 \\
CO & 4.0~10$^{-8a}$ & 6~10$^{-8}$ & 0.667 \\
CO$_{2}$ & 8.5~10$^{-7}$ & 6.2~10$^{-7}$ & 1.371 \\
\hline
\end{tabular}
 \tablefoot{$^{a}$ This rate comes from \cite{Huebner1979}.}
\end{table}

\section{Observations}

We carried out the observations on four comets i.e C/2002 T7 (LINEAR), 73P-C/Schwassmann-Wachmann 3, 8P/Tuttle, and 103P/Hartley 2 (hereafter C/2002 T7, 73P-C, 8P and 103P, respectively), using the high resolution UVES spectrograph at ESO (VLT). The observations of comet C/2002 T7  have been done in May 2004 shortly after its perihelion (April 2004).
In 1995, comet 73P/Schwassmann-Wachmann~3 was very active and the nucleus split into several fragments \citep{Bohnhardt1995}. The oxygen emission lines have been observed on the brightest fragment which is 73P-C.
Spectra of comet 8P were obtained on three different nights during January and February 2008. 
The Jupiter family comet 103P/Hartley~2 was studied in great details by the EPOXI NASA mission and many telescopes from September to November 2010 \citep{AHearn2011, Meech2011}. Our observations were scheduled at the Paranal Observatory during the EPOXI flyby period. \\
All the data correspond to spectroscopic observations made with the high-resolution UVES spectrograph mounted on the UT2 of the VLT \citep[and references therein]{Manfroid2009, Jehin2009, Decock2013}. The peculiarity of these observations is that these comets were observed close to the Earth, from 0.61 au for C/2002 T7 down to 0.15 au for 73P-C. These small geocentric distances provide a high spatial resolution (from 400 to 100 km per arc second) and permit us to analyze the coma species very close to the nucleus. The complete sample is made of five spectra for C/2002 T7, eight for 103P, three for 8P and one for 73P-C. The 0.4$^{\prime \prime}\times$12$^{\prime\prime}$ slit is usually centered on the nucleus. In the case of C/2002 T7 and 103P, we also obtained spectra at offset positions making it possible to study the [OI] lines at large distances from the nucleus. All the observing details as well as the offset values are given in Table~\ref{observations}.

\begin{table*}[H]
\centering
\scriptsize
\label{observations}
\caption{Observational circumstances.} 
\begin{tabular}{l l l l l l l l l l l l}
\hline
\hline
Comet & UT Date & $r$ & $\dot{r}$ & $\Delta$ & $\dot{\Delta}$ & Exptime & Offset  & Slit & Slit  & Seeing & \hfill{Seeing}\\
& & (au) & (km~s$^{-1}$) & (au) & (km~s$^{-1}$) & (s) & ($^{\prime \prime}$) & ($^{\prime \prime}\times^{\prime\prime}$) & (km~$\times$~km) & ($^{\prime\prime}$) & \hfill{(km)} \\
\hline
C/2002 T7 (LINEAR) & 2004/05/06, 10:15 & 0.68 & 15.83 & 0.61 & -65.62 & 1080 & 5 & 0.44 $\times$ 12.00 & 195 $\times$ 5~309 & 1.0 & \hfill{458} \\
C/2002 T7 (LINEAR) & 2004/05/26, 23:47 & 0.94 & 25.58 & 0.41 & 54.98 & 2678 & 0 &  0.30 $\times$ 12.00 & 89 $\times$ 3~568 & 0.7 & \hfill{198} \\
C/2002 T7 (LINEAR) & 2004/05/27, 01:06 & 0.94 & 25.59 & 0.42 & 55.20 & 1800 & 0 &  0.30 $\times$ 12.00 & 91 $\times$ 3~655 & 0.7 & \hfill{206} \\
C/2002 T7 (LINEAR) & 2004/05/28, 23:24 & 0.97 & 25.90 & 0.48 & 59.15 & 487 & 70 &  0.44 $\times$ 12.00 & 153 $\times$ 4~178 & 1.0 & \hfill{354} \\
C/2002 T7 (LINEAR) & 2004/05/29, 00:09 & 0.97 & 25.90 & 0.48 & 59.26 & 3600 & 70 &  0.44 $\times$ 12.00 & 153 $\times$ 4~178 & 1.0 & \hfill{205} \\
73P-C SW 3 & 2006/05/27, 09:28 & 0.97 & -4.17 & 0.15 & 12.31 & 4800 & 0 & 0.60 $\times$ 12.00 & 65 $\times$ 2~611 & 1.4 & \hfill{151} \\
8P/Tuttle & 2008/01/16, 01:00 & 1.04 & -4.29 & 0.36 & 21.64 & 3600 & 0 & 0.44 $\times$ 10.00 & 115 $\times$ 1~305 & 1.0 & \hfill{261} \\
8P/Tuttle & 2008/01/28, 00:58 & 1.03 & 0.40 & 0.52 & 24.72 & 3900 & 0 & 0.44 $\times$ 10.00 & 166 $\times$ 3~771 & 0.9 & \hfill{319} \\
8P/Tuttle & 2008/02/04, 00:57 & 1.03 & 3.16 & 0.62 & 24.16 & 3900 & 0 & 0.44 $\times$ 10.00 & 198 $\times$ 4~497 & 1.0 & \hfill{450} \\
103P/Hartley 2 & 2010/11/05, 07:18 & 1.06 & 2.53 & 0.16 & 7.08 & 2900 & 0 & 0.44 $\times$ 12.00 & 51 $\times$ 1~393 & 1.3 & \hfill{148} \\
103P/Hartley 2 & 2010/11/05, 08:19 & 1.06 & 2.55 & 0.16 & 7.19 & 3200 & 0 & 0.44 $\times$ 12.00 & 57 $\times$ 1~567 & 1.8 & \hfill{205} \\
103P/Hartley 2 & 2010/11/10, 07:17 & 1.07 & 4.05 & 0.18 & 7.96 & 2900 & 0 & 0.44 $\times$ 12.00 & 57 $\times$ 1~567 & 0.8 & \hfill{102} \\
103P/Hartley 2 & 2010/11/10, 08:19 & 1.07 & 4.07 & 0.18 & 8.07 & 3200 & 0 & 0.44 $\times$ 12.00 & 51 $\times$ 1~393 & 0.8 & \hfill{97} \\
103P/Hartley 2 & 2010/11/11, 05:53 & 1.08 & 4.33 & 0.19 & 7.94 & 4500 & 30 & 0.44 $\times$ 12.00 & 60 $\times$ 1~654 & 0.6 & \hfill{76} \\
103P/Hartley 2 & 2010/11/11, 07:18 & 1.08 & 4.35 & 0.19 & 8.08 & 3600 & 20 & 0.44 $\times$ 12.00 & 60 $\times$ 1~654 & 0.6 & \hfill{81} \\
103P/Hartley 2 & 2010/11/11, 08:16 & 1.08 & 4.36 & 0.19 & 8.19 & 2400 & 10 & 0.44 $\times$ 12.00 & 60 $\times$ 1~654 & 0.6 & \hfill{82} \\
103P/Hartley 2 & 2010/11/11, 08:57 & 1.08 & 4.37 & 0.19 & 8.27 & 900 & 0 & 0.44 $\times$ 12.00 & 60 $\times$ 1~654 & 0.7 & \hfill{92} \\
\hline
\hline
\end{tabular}
\tablefoot{$r$ is the heliocentric distance, $\Delta$ is the geocentric distance. $\dot{r}$ and $\dot{\Delta}$ are, respectively, the heliocentric and geocentric velocities. Exptime corresponds to the exposure time in seconds. Offset (arc seconds) corresponds to the distance between the slit and the center of the cometary nucleus. The slit-tail orientation provides the direction of the slit with respect to the cometary tail.
Slit is the size of the entrance slit of the spectrograph in arc seconds. The next column gives the projected area covered by the slit. The last two columns present the seeing values in arc seconds and in km during the observations.}
\end{table*}

\normalsize

\section{Data reduction}

\begin{figure}[h!]
\centerline{\includegraphics[width=\columnwidth]{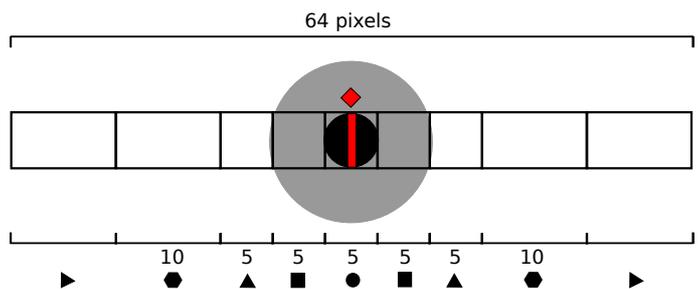}}
\caption{A representation of the slit sub-division. Different symbols for different pixel zones are used in Figures~\ref{gr-linear} to \ref{gr-hart}. The numbers given above the symbols correspond to the size of the sub-slit in pixels. Note that the central pixel is represented by the red diamond symbol above the slit.}
\label{sub_slit}
\end{figure}

The extraction of the 1D spectra and the reduction procedure are described in Sect.~3 of \cite{Decock2013}. In order to study the [OI] lines as close as possible to the comet nucleus, we spatially extracted the spectra line by line. We divided the slit of each spectrum into different pixel zones corresponding to different distances to the nucleus. As shown in Fig.~\ref{sub_slit}, various sub-slit zones are represented by different symbols and will be used in Figures~\ref{gr-linear} to \ref{gr-hart}. We obtained 14 spectra for comets C/2002 T7, 73P-C and 103P along the 64 pixels slit whereas for comet 8P, five spectra are extracted for the 52 pixels slit length. For the spectra taken at large offset positions, we use the average value over the whole slit to increase the signal-to-noise ratio, because in this case the change of distance from the nucleus is small within the slit.

\subsection{C$_{2}$ emission lines subtraction}

\begin{figure}[h!]
\centerline{\includegraphics[width=\columnwidth]{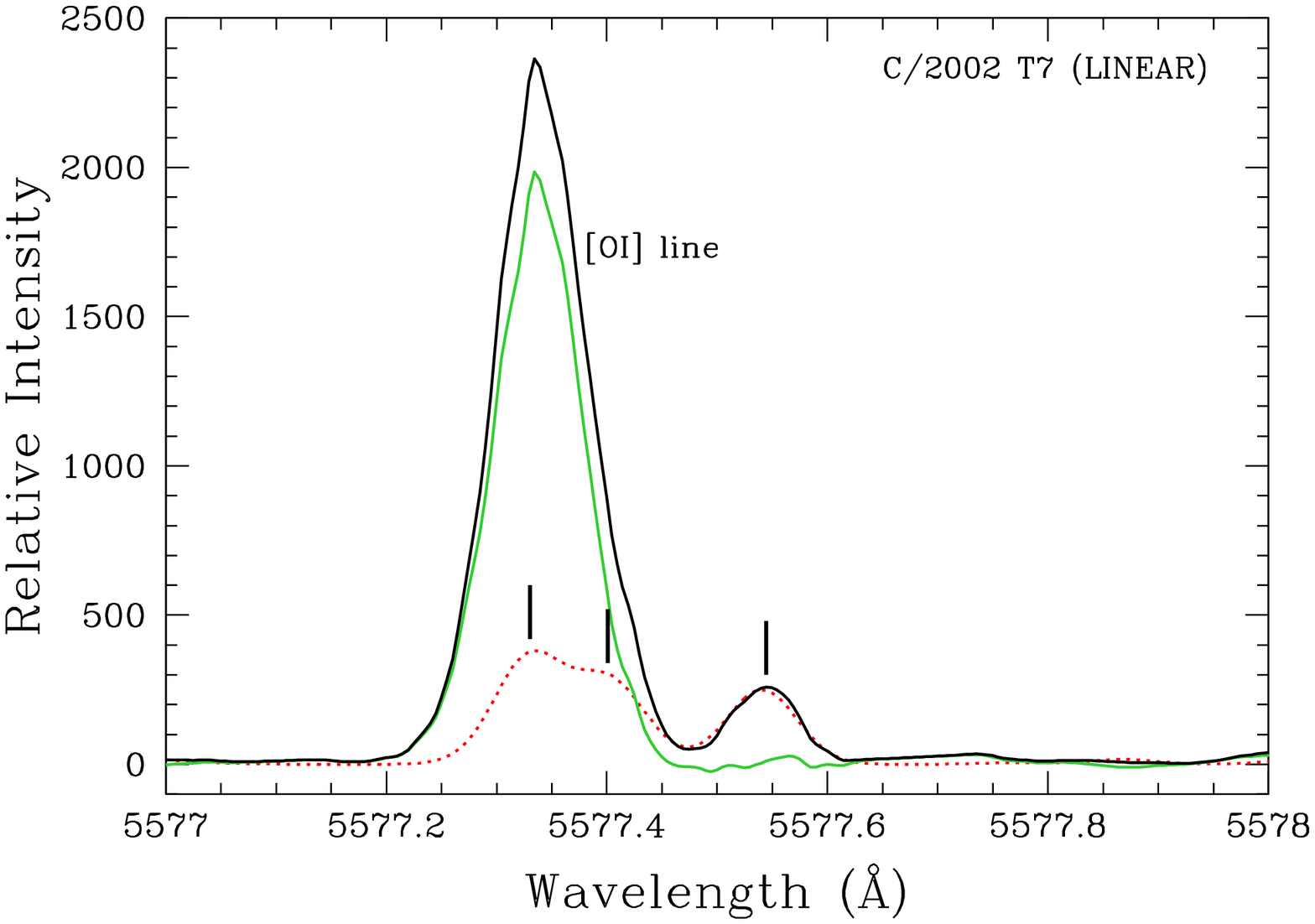}}
\centerline{\includegraphics[width=\columnwidth]{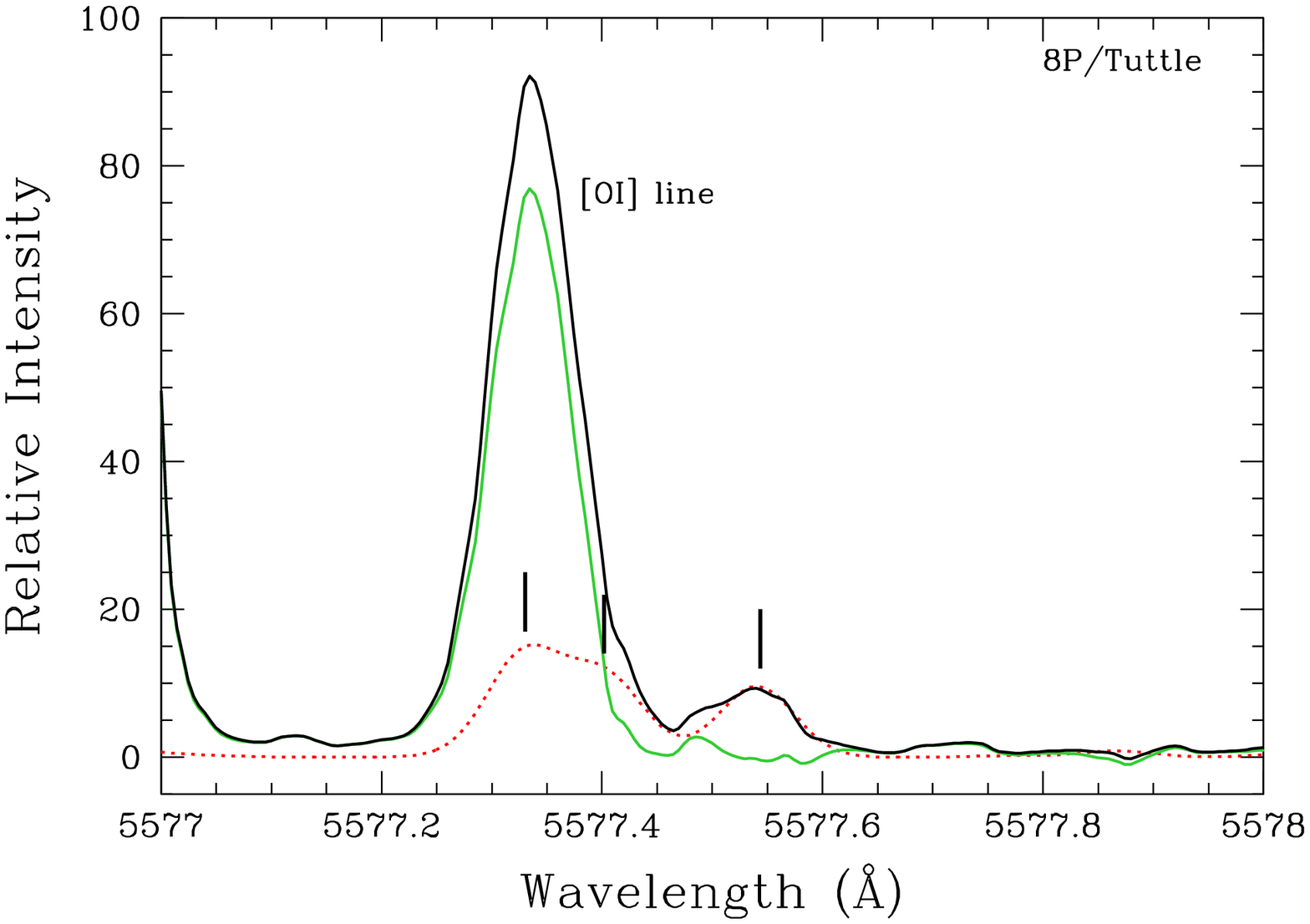}}
\centerline{\includegraphics[width=\columnwidth]{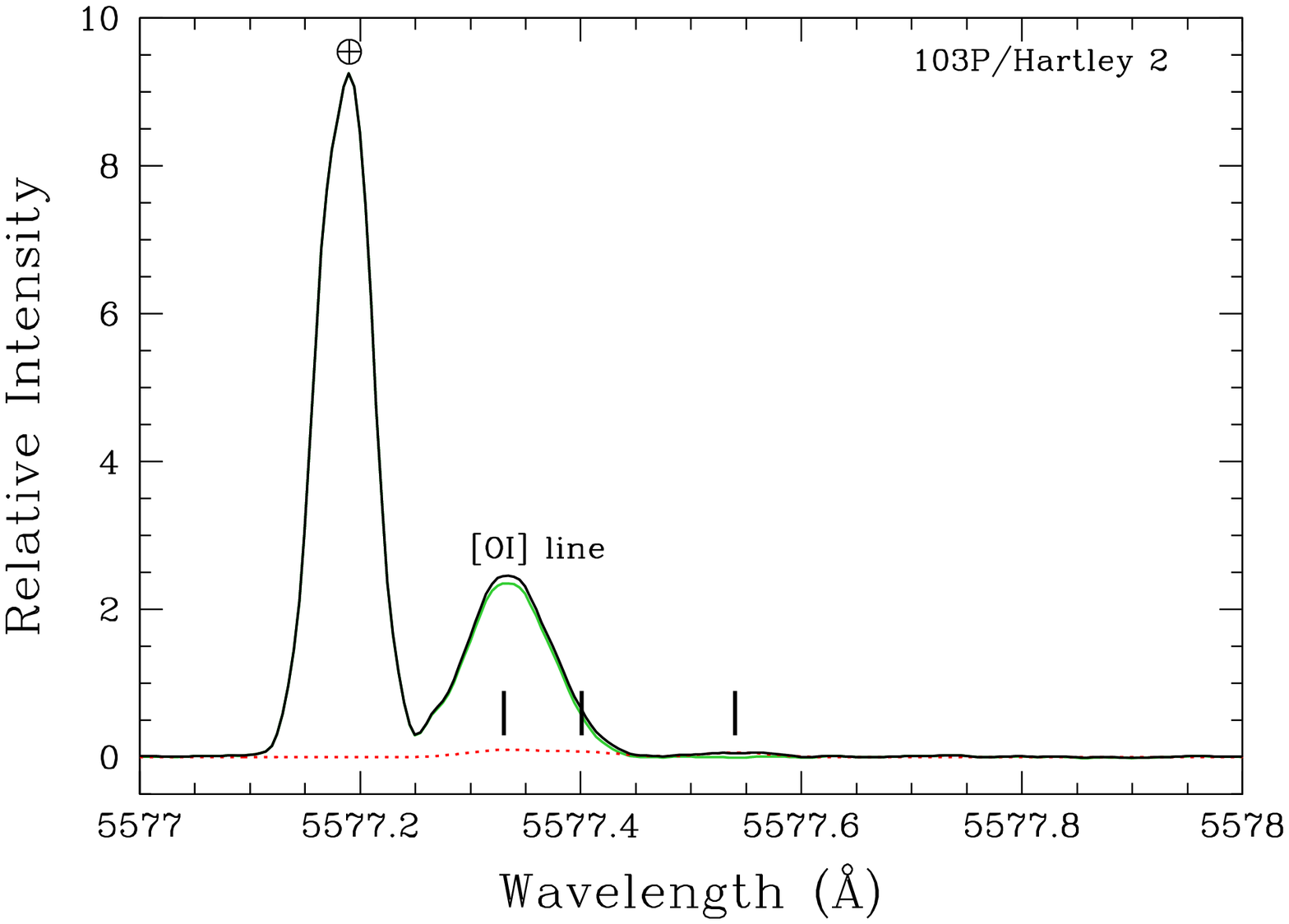}}
\caption{Subtraction of C$_{2}$ lines from the spectra of comets C/2002 T7 (LINEAR), 8P/Tuttle and 103P/Hartley 2. The black spectrum shows the data not corrected for C$_{2}$ while in the green one is after removing the C$_{2}$ lines contamination. The C$_{2}$ synthetic spectrum is represented by the dotted red line. The positions of C$_{2}$ lines at 5577.331 $\AA$, 5577.401 $\AA$ and 5577.541 $\AA$ are indicated with small vertical thick marks. The telluric [OI] line ($\oplus$) is visible in the spectrum of comet 103P/Hartley~2.}
\label{soustraction_c2}
\end{figure}

Many C$_{2}$ lines are located in the vicinity of the green line and we have identified two of them at 5577.331 $\AA$ and 5577.401 $\AA$ which are blended with the oxygen line. 
The C$_{2}$ contamination becomes more important at large distances from the nucleus as the [OI] flux decreases faster than C$_{2}$. This blend can affect both the flux and the width of the green line.
In order to remove this contamination, we used another C$_2$ line close to the oxygen line (the P$_3(25)$ line of
the (1,2) Swan band at 5577.541~$\AA$) and created a synthetic C$_2$ spectrum
by fitting both the intensity and 
the profile of the lines. This synthetic spectrum was built using a Boltzman distribution of the rotational levels relative populations with a temperature of 4000~K. Because of the very close energy levels
involved (same rotational number value for both the upper and lower vibronic states, see 
\citealt{Rousselot2012} for more details of the Swan bands structure) the exact
rotational temperature value has very little influence on the final C$_2$
fit. The spectrum considered in our work is the observed one
after subtraction of the C$_2$ synthetic spectrum (see Fig.~\ref{soustraction_c2}). This step was not needed for 73P-C because there is no C$_{2}$ feature detected around the green line, this comet being a C-chain depleted comet \citep{Schleicher2006} and thus very poor in C$_{2}$.

\section{Results and discussion}

We used the IRAF\footnote{IRAF is a tool for the reduction and the analysis of astronomical data (http://iraf.noao.edu).} software to measure the intensities and the widths of the three [OI] lines, by making a gaussian fit.

\subsection{G/R ratio}

\onllongtab{
\begin{landscape}
\begin{table}[h!]
\scriptsize
\begin{tabular}{l c c c c c c c c c c c r}
\hline
\hline
Comet & UT Date & Nucleocentric projected & \multicolumn{3}{c}{Intensity (ADU)} & G/R & \multicolumn{3}{c}{FWHM$_{\rm{observed}}$~($\AA$)} & \multicolumn{3}{c}{FWHM$_{\rm{intrinsic}}$~(km~s$^{-1}$)}  \\ 
 & & distance range (km) & 5577.339~\r{A} & 6300.304~\r{A} & 6363.776~\r{A} & & 5577.339~\r{A} & 6300.304~\r{A} & 6363.776~\r{A} & 5577.339~\r{A} & 6300.304~\r{A} &  \hfill{6363.776~\r{A}} \\
 \hline
C/2002 T7 (LINEAR) & 2004/05/06, 10:15 & 2212 & 3088 & 1015 & 201 & 0.049 & 0.095 & 0.099 & 0.100 & 2.329  & 1.819 & 1.795 \\
 & 2004/05/26, 23:47 & 0 - 28 & 1899 & 569 & 426 & 0.172 & 0.096 & 0.090 & 0.091 & 2.634 & 1.759 & 1.734 \\
 &  & 0 - 139 & 1394 & 425 & 331 & 0.182 & 0.097 & 0.091 & 0.092 & 2.691 & 1.780 & 1.796 \\
 &  & 139 - 418 & 1576 & 490 & 317 & 0.154 & 0.099 & 0.090 & 0.093 & 2.747 & 1.734 & 1.836 \\
 &  & 418 - 697 & 1842 & 560 & 203 & 0.084 & 0.097 & 0.089 & 0.090 & 2.676 & 1.717 & 1.696 \\
 &  & 697 - 976 & 1523 & 456 & 139 & 0.070 & 0.095 & 0.088 & 0.090 & 2.623 & 1.644 & 1.713 \\
 &  & 976 - 1784 & 1155 & 358 & 84 & 0.056 & 0.094 & 0.089 & 0.090 & 2.554 & 1.683 & 1.729 \\
 & 2004/05/27, 01:06 & 0 - 29 & 1837 & 555 & 289 & 0.121 & 0.097 & 0.089 & 0.091 & 2.695 & 1.700 & 1.771 \\
 &  & 0 - 143 & 1690 & 515 & 280 & 0.127 & 0.100 & 0.090 & 0.092 & 2.792 & 1.738 & 1.775 \\
 &  & 143 - 428 & 1675 & 503 & 245 & 0.114 & 0.093 & 0.088 & 0.095 & 2.543 & 1.644 & 1.913 \\
 &  & 428 - 714 & 1763 & 533 & 171 & 0.074 & 0.095 & 0.089 & 0.090 & 2.623 & 1.687 & 1.717 \\
 &  & 714 - 1000 & 1504 & 450 & 120 & 0.062 & 0.094 & 0.087 & 0.091 & 2.554 & 1.595 & 1.750 \\
 &  & 1000 - 1828 & 1139 & 345 & 77 & 0.052 & 0.091 & 0.089 & 0.091 & 2.470 & 1.683 & 1.738 \\
 & 2004/05/28, 23:24 & 24522 & 47 & 14 & 4 & 0.065 & 0.088 & 0.100 & 0.096 & 2.043 & 1.822 & 1.610 \\
 & 2004/05/29, 00:09 & 24522 & 40 & 12 & 4 & 0.071 & 0.084 & 0.101 & 0.100 & 1.868 & 1.870 & 1.758 \\
 73P-C/SW 3 & 2006/05/27, 09:31 & 0 - 10 & 712 & 234 & 189 & 0.200 & 0.107 & 0.109 & 0.108 & 2.278 & 1.493 & 1.369 \\
 & & 0 - 51 & 697 & 232 & 188 & 0.202 & 0.107 & 0.108 & 0.108 & 2.253 & 1.475 & 1.419 \\
 & & 51 - 153 & 612 & 202 & 145 & 0.178 & 0.106 & 0.107 & 0.111 & 2.214 & 1.421 & 1.594 \\
 & & 153 -  255 & 595 & 203 & 92 & 0.115 & 0.106 & 0.109 & 0.107 & 2.204 & 1.493 & 1.312 \\
 & & 255 - 357 & 519 & 174 & 64 & 0.092 & 0.105 & 0.107 & 0.105 & 2.143 & 1.396 & 1.211 \\
 & & 357 - 653 & 405 & 133 & 18 & 0.033 & 0.103 & 0.107 & 0.106 & 2.077 & 1.390 & 1.266 \\
8P/Tuttle & 2008/01/16, 01:00 & 0 - 25 & 182 & 61 & 15 & 0.063 & 0.088 & 0.092 & 0.093 & 2.030 & 1.481 & 1.477 \\
 & & 0 - 126 & 172 & 59 & 16 & 0.068 & 0.090 & 0.093 & 0.093 & 2.084 & 1.575 & 1.482 \\
 & & 126 - 377 & 162 & 53 & 11 & 0.052 & 0.090 & 0.093 & 0.094 & 2.106 & 1.551 & 1.517 \\
 & & 377 - 628 & 144 & 47 & 6 & 0.034 & 0.083 & 0.092 & 0.091 & 1.762 & 1.501 & 1.363 \\
 & & 628 - 1305 & 102 & 34 & 4 & 0.028 & 0.089 & 0.092 & 0.091 & 1.858 & 1.491 & 1.411 \\
 & 2008/01/28, 00:58 & 0 - 36 & 547 & 185 & 68 & 0.093 & 0.089 & 0.094 & 0.094 & 2.068 & 1.584 & 1.544 \\
 & & 0 - 181 & 475 & 155 & 59 & 0.094 & 0.091 & 0.093 & 0.095 & 2.122 & 1.520 & 1.597 \\
 & & 181 - 544 & 558 & 185 & 25 & 0.034 & 0.089 & 0.088 & 0.093 & 2.046 & 1.261 & 1.459 \\
 & & 544 - 907 & 305 & 100 & 12 & 0.029 & 0.088 & 0.093 & 0.091 & 2.014 & 1.535 & 1.387 \\
 & & 907 - 1886 & 182 & 60 & 7 & 0.027 & 0.085 & 0.093 & 0.092 & 1.884 & 1.530 & 1.403 \\
 & 2008/02/04, 00:57 & 0 - 43 & 452 & 149 & 36 & 0.059 & 0.090 & 0.093 & 0.091 & 2.093 & 1.477 & 1.332 \\
 & & 0 - 216 & 408 & 135 & 32 & 0.060 & 0.088 & 0.092 & 0.092 & 1.974 & 1.462 & 1.396 \\
 & & 216 - 649 & 465 & 130 & 23 & 0.039 & 0.089 & 0.088 & 0.091 & 2.038 & 1.238 & 1.348 \\
 & & 649 - 1081 & 311 & 101 & 14 & 0.034 & 0.089 & 0.092 & 0.091 & 2.029 & 1.420 & 1.310 \\
 & & 1081 - 2248 & 193 & 64 & 9 & 0.033 & 0.091 & 0.093 & 0.091 & 2.102 & 1.493 & 1.337 \\
 103P/Hartley 2 & 2010/11/05, 07:18 & 0 - 11 & 2.3 & 0.7 & 0.7 & 0.224 & 0.094 & 0.084 & 0.082 & 2.395 & 1.388 & 1.208 \\
 & & 0 - 54 & 2.1 & 0.7 & 0.7 & 0.240 & 0.095 & 0.084 & 0.086 & 2.432 & 1.413 & 1.428 \\
 & & 54 - 163 & 2.2 & 0.7 & 0.6 & 0.190 & 0.091 & 0.087 & 0.086 & 2.241 & 1.513 & 1.389 \\
 & & 163 - 272 & 2.0 & 0.6 & 0.4 & 0.138 & 0.092 & 0.087 & 0.081 & 2.292 & 1.518 & 1.164 \\
 & & 272 - 381 & 1.6 & 0.5 & 0.2 & 0.112 & 0.088 &  0.084 & 0.083 & 2.109 & 1.403 & 1.256 \\
 & & 381 - 696 & 1.3 & 0.4 & 0.2 & 0.094 & 0.084 & 0.083 & 0.084 & 1.955 & 1.328 & 1.324 \\
 & 2010/11/05, 08:19 & 0 - 12 & 2.4 & 0.8 & 0.8 & 0.259 & 0.094 & 0.084 & 0.085 & 2.387 & 1.398 & 1.339 \\
 & & 0 - 61 & 2.2 & 0.7 & 0.8 & 0.292 & 0.097 & 0.087 & 0.087 & 2.505 & 1.536 & 1.457 \\
 & & 61 - 184 & 2.2 & 0.7 & 0.7 & 0.236 & 0.097 & 0.084 & 0.086 & 2.505 & 1.413 & 1.389 \\
 & & 184 - 306 & 2.0 & 0.7 & 0.4 & 0.167 & 0.092 & 0.087 & 0.080 & 2.308 & 1.518 & 1.119 \\
 & & 306 - 428 & 1.8 & 0.6 & 0.3 & 0.121 & 0.093 & 0.083 & 0.084 & 2.337 & 1.349 & 1.288 \\
 & & 428 - 783 & 1.4 & 0.4 & 0.2 & 0.087 & 0.085 & 0.083 & 0.084 & 1.991& 1.349 & 1.314 \\
 & 2010/11/10, 07:17 & 0 - 11 & 2.8 & 0.9 & 0.9 & 0.256 & 0.093 & 0.082 & 0.084 & 2.353 & 1.316 & 1.354 \\
 & & 0 - 54 & 2.4 & 0.8 & 0.9 & 0.288 & 0.092 & 0.084 & 0.083 & 2.312 & 1.396 & 1.267 \\
 & & 54 - 163 & 2.4 & 0.6 & 0.5 & 0.180 & 0.093 & 0.081 & - & 2.329 & 1.249 & -  \\
 & & 163 - 272 & 2.1 & 0.7 & 0.3 & 0.096 & 0.087 & 0.078 & 0.082 & 2.113 & 1.070 & 1.209 \\
 & & 272 - 381 & 1.7 & 0.5 & 0.2 & 0.081 & 0.087 & 0.081 & 0.083 & 2.109 & 1.249 & 1.303 \\
 & & 381 - 696 & 1.2 & 0.4 & 0.1 & 0.067 & 0.083 & 0.082 & 0.082 & 1.915 & 1.286 & 1.241 \\
 & 2010/11/10, 08:19 & 0 - 12 & 2.8 & 0.9 & 0.8 & 0.211 & 0.092 & 0.081 & 0.082 & 2.304 & 1.222 & 1.252 \\
 & & 0 - 61 & 2.0 & 0.7 & 0.7 & 0.272 & 0.094 & 0.082 & 0.082 & 2.407 & 1.280 & 1.241 \\
 & & 61 - 184 & 2.3 & 0.7 & 0.6 & 0.191 & 0.091 & 0.079 & 0.082 & 2.270 & 1.111 & 1.241 \\
 & & 184 - 306 & 2.3 & 0.7 & 0.3 & 0.111 & 0.088 & 0.081 & 0.080 & 2.148 & 1.270 & 1.097 \\
 & & 306 - 428 & 1.8 & 0.6 & 0.2 & 0.099 & 0.091 & 0.083 & 0.083 & 2.262 & 1.352 & 1.262 \\
 & & 428 - 783 & 1.3 & 0.4 & 0.2 & 0.088 & 0.087 & 0.082 & 0.083 & 2.083 & 1.311 & 1.278 \\
 & 2010/11/11, 05:53 & 4134 & 0.2 & 0.1 & 0.02 & 0.064 & 0.079 & 0.083 & 0.083 & 1.739 & 1.338 & 1.269 \\
 & 2010/11/11, 07:18 & 2756 & 0.8 & 0.3 & 0.04 & 0.046 & 0.081 & 0.082 & 0.082 & 1.833 & 1.300 & 1.195 \\
 & 2010/11/11, 08:16 & 1378 & 0.3 & 0.1 & 0.02 & 0.042 & 0.081 & 0.081 & 0.081 & 1.823 & 1.228 & 1.183 \\
 & 2010/11/11, 08:57 & 0 - 13 & 2.4 & 0.8 & 1.2 & 0.359 & 0.094 & 0.083 & 0.085 & 2.372 & 1.327 & 1.385 \\
 & & 0 - 65 & 2.0 & 0.7 & 1.1 & 0.417 & 0.096 & 0.083 & 0.090 & 2.442 & 1.327 & 1.600 \\
 & & 65 - 194 & 2.6 & 0.8 & 0.7 & 0.195 & 0.095 & 0.079 & 0.078 & 2.409 & 1.111 & 1.014 \\
 & & 194 - 323 & 1.8 & 0.6 & 0.3 & 0.135 & 0.095 & 0.081 & 0.084 & 2.401 & 1.244 & 1.320 \\
 & & 323 - 452 & 1.5 & 0.5 & 0.2 & 0.108 & 0.088 & 0.082 & 0.082 & 2.141 & 1.286 & 1.204 \\
 & & 452 - 827 & 1.1 & 0.4 & 0.1 & 0.089 & 0.088 & 0.082 & 0.082 & 2.137 & 1.296 & 1.252 \\
\hline
\end{tabular}
\tablefoot{For the offset spectra, we give the distance to the center of the slit. For the third subslit spectrum of the third observation of comet 103P/Hartley 2, the 6363.776 $\AA$ line could not be used because of the contamination by a strong cosmic ray event in that pixel zone. The offset spectra are indicated with an asterisk.} \label{tableaulong} 
\end{table}
\end{landscape}
\normalsize
}

\onllongtab{
\begin{table}[h!]
\scriptsize
\caption{Average values of the G/R ratio and the intrinsic line FWHM for each spatial bin.}

\begin{tabular}{l c c c c c c r}
\hline
\hline
Comet & UT Date & N & Average nucleocentric & G/R & \multicolumn{3}{c}{FWHM$_{\rm{intrinsic}}$(km~s$^{-1}$)}  \\ 
 & & & distance range (km) & & 5577.339 $\AA$ & 6300.304 $\AA$ & 6363.776 $\AA$ \\ 
 \hline
\label{average}
C/2002 T7 (LINEAR) & 2004/05/06$^{*}$ & 1 & 2212 & 0.049 & 2.329 & 1.819 & 1.795 \\
 & 2004/05/26-27 & 2 & 0 - 29 & 0.147 $\pm$ 0.036 & 2.664 $\pm$ 0.043 & 1.730 $\pm$ 0.042 & 1.752 $\pm$ 0.026 \\
 &  & 2 & 0 - 141 & 0.155 $\pm$ 0.039 & 2.741 $\pm$ 0.072 & 1.759 $\pm$ 0.030 & 1.785 $\pm$ 0.015 \\
 &  & 2 & 141 - 423 & 0.134 $\pm$ 0.028 & 2.645 $\pm$ 0.144 & 1.689 $\pm$ 0.064 & 1.875 $\pm$ 0.054 \\
 &  & 2 & 423 - 706 & 0.079 $\pm$ 0.007 & 2.649 $\pm$ 0.037 & 1.702 $\pm$ 0.021 & 1.706 $\pm$ 0.015 \\
 &  & 2 & 706 - 988 & 0.066 $\pm$ 0.006 & 2.589 $\pm$ 0.048 & 1.620 $\pm$ 0.034 & 1.731 $\pm$ 0.027  \\
 &  & 2 & 988 - 1806 & 0.054 $\pm$ 0.003 & 2.512 $\pm$ 0.060 & 1.683 $\pm$ 0.000 & 1.734 $\pm$ 0.006 \\
 & 2004/05/28-29$^{*}$ & 2 & 24522 & 0.068 $\pm$ 0.004 & 1.955 $\pm$ 0.123 & 1.846 $\pm$ 0.034 & 1.684 $\pm$ 0.105 \\
 73P-C/SW 3 & 2006/05/27 & 1 & 0 - 10 & 0.200 & 2.278 & 1.493 & 1.369 \\
 & & 1 & 0 - 51 & 0.202 & 2.253 & 1.475 & 1.419 \\
 & & 1 & 51 - 153 & 0.178 & 2.214 & 1.421 & 1.594 \\
 & & 1 & 153 - 255 & 0.115 & 2.204 & 1.493 & 1.312 \\
 & & 1 & 255 - 357 & 0.092 & 2.143 & 1.396 & 1.211 \\
 & & 1 & 357 - 653 & 0.033 & 2.077 & 1.390 & 1.266 \\
8P/Tuttle & 2008/01/16-28, 02/04 & 3 & 0 - 35 & 0.072 $\pm$ 0.019 & 2.064 $\pm$ 0.032 & 1.514 $\pm$ 0.060 & 1.451 $\pm$ 0.109 \\
 & & 3 & 0 - 174 & 0.074 $\pm$ 0.018 & 2.060 $\pm$ 0.077 & 1.519 $\pm$ 0.057 & 1.492 $\pm$ 0.101 \\
 & & 3 & 174 - 523 & 0.042 $\pm$ 0.009 & 2.063 $\pm$ 0.037 & 1.350 $\pm$ 0.175 & 1.442 $\pm$ 0.086 \\
 & & 3 & 523 - 872 & 0.032 $\pm$ 0.003 & 1.935 $\pm$ 0.150 & 1.485 $\pm$ 0.059 & 1.353 $\pm$ 0.040 \\
 & & 3 & 872 - 1813 & 0.030 $\pm$ 0.003 & 1.948 $\pm$ 0.134 & 1.505 $\pm$ 0.022 & 1.383 $\pm$ 0.040 \\
 103P/Hartley 2 & 2010/11/05-10-11 & 5 & 0 - 12 & 0.262 $\pm$ 0.058 & 2.362 $\pm$ 0.036 & 1.330 $\pm$ 0.071 & 1.307 $\pm$ 0.074 \\
 & & 5 & 0 - 59 & 0.302 $\pm$ 0.068 & 2.420 $\pm$ 0.070 & 1.276 $\pm$ 0.047 & 1.187 $\pm$ 0.089 \\
 & & 5 & 59 - 178 & 0.198 $\pm$ 0.022 & 2.351 $\pm$ 0.107 & 1.279 $\pm$ 0.180 & 1.258 $\pm$ 0.177 \\
 & & 5 & 178 - 296 & 0.129 $\pm$ 0.027 & 2.252 $\pm$ 0.119 & 1.324 $\pm$ 0.193 & 1.182 $\pm$ 0.088 \\
 & & 5 & 296 - 414 & 0.104 $\pm$ 0.015 & 2.192 $\pm$ 0.103 & 1.328 $\pm$ 0.061 & 1.263 $\pm$ 0.038 \\
 & & 5 & 414 - 757 & 0.085 $\pm$ 0.010 & 2.016 $\pm$ 0.092 & 1.314 $\pm$ 0.025 & 1.282 $\pm$ 0.037 \\
 & 2010/11/11, 05:53$^{*}$ & 1 & 4134 & 0.064 & 1.739 & 1.338 & 1.269 \\
 & 2010/11/11, 07:18$^{*}$ & 1 & 2756 & 0.046 & 1.833 & 1.300 & 1.195 \\
 & 2010/11/11, 08:16$^{*}$ & 1 & 1378 & 0.042 & 1.825 & 1.228 & 1.183 \\
\hline
\hline
\end{tabular}
\tablefoot{The errors listed are the rms of the N spectra available.}
\end{table}
\normalsize
}

\begin{figure}[h!]
\centerline{\includegraphics[width=\columnwidth]{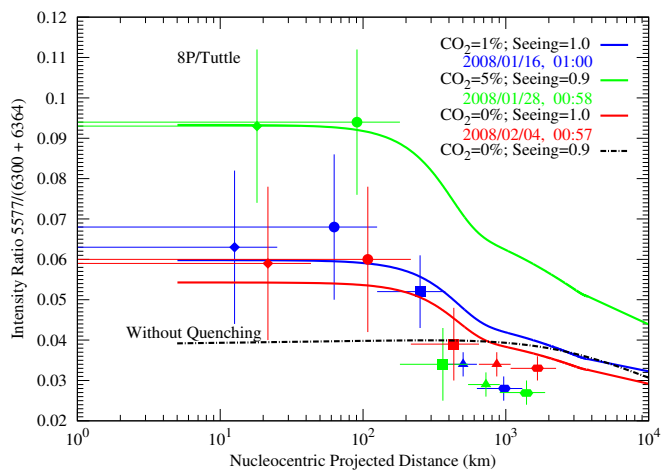}}
\caption{The G/R ratio for each sub-slit spectra for comet 8P/Tuttle. The range of nucleocentric distances covered by each point as well as the error on the G/R ratio are represented. The model calculated green to red-doublet emission intensity ratios as a function of projected distance
in comet are plotted. A H$_2$O production rate of 1.4 $\times$ 10$^{28}$ s$^{-1}$ 
 and 0.5\% CO as relative abundance have been used with different seeing values and with different CO$_{2}$ relative abundances. The fits give the CO$_{2}$ relative abundance of the comet. The black dash-dotted curve represents the calculated G/R ratio accounting for no collisional quenching with 0$\%$ CO$_{2}$ relative abundance.}
\label{gr-tuttle}
\end{figure}

When collisional quenching is neglected, the intensity $I$ of an emission line, expressed in Rayleighs\footnote{1R =~10$^{6}$~photons cm$^{-2}$ s$^{-1}$~\textnormal{in 4$\pi$ steradians}} (R), is given theoretically by \cite{Festou1981} :
\begin{eqnarray}
\label{intensity}
I = ~~ 10^{-6} ~~ \tau_{p}^{-1} ~~ \alpha ~~ \beta ~~ N 
\end{eqnarray}
where $\tau_{p}$ is the photodissociative lifetime of the parent species in seconds, $\alpha$ is the yield of photodissociation, $\beta$ corresponds to the branching ratio for the transition, and $N$ is the column density of the parent species in cm$^{-2}$. The measured green and red-doublet emission intensities and the corresponding G/R ratios for each spectrum of our observations are listed in Table~\ref{tableaulong}. 
The G/R ratio is displayed for the four comets with respect to the nucleus distance in Figures~\ref{gr-linear}, \ref{gr-sw}, \ref{gr-tuttle} and \ref{gr-hart}. It is given in Table~\ref{average} with the errors on the y-axis estimated from the rms of the N spectra available for a given comet. Errors on the x-axis correspond to the spatial area (in km) covered by the considered sub-slit. The atmospheric seeing limits the spatial resolution and defines a minimal nucleocentric distance that could be resolved. This minimal nucleocentric distance is provided in the last column of Table~\ref{observations} and is also included in the x-axis errors. Referring to the latter figures, we notice that the G/R ratio decreases monotonically with the projected distance from the nucleus. This is due to strong collisional quenching of the excited oxygen atoms by H$_{2}$O \citep{Bhardwaj2012}. Since the lifetime of O($^{1}$D) atoms is $\sim$100 times longer than the one of O($^{1}$S), the destruction of O($^{1}$D) by quenching is more important \citep{Raghuram2014} and implies an increase of the G/R ratio close to the nucleus as we see in our observations. Without accounting for collisional quenching of O($^{1}$S) and O($^{1}$D), the model calculated G/R ratio profiles are plotted with dash-dotted curves in Figures~\ref{gr-linear} to \ref{gr-hart}. The G/R ratio is almost flat throughout the projected coma of all these comets and is significantly smaller compared to the observed values very close to the nucleus. This calculation shows the importance of the collisional quenching in determining the G/R ratio in comets. The destruction rate profiles of O($^{1}$S) and O($^{1}$D) by H$_{2}$O in the innermost coma depends on the size of the collisional coma of the comet which is proportional to the water production rate. The larger the H$_{2}$O production rate, the more extended the collisional coma and the stronger the quenching by H$_{2}$O molecules. That is why comet C/2002 T7 (LINEAR), with a higher Q$_{\rm{H_{2}O}}$ $\sim$~5~$\times$~10$^{29}$~s$^{-1}$, reaches the minimal value of the G/R ratio at larger distances ($\sim$1000~km) than to the other comets ($\sim$500~km). \\
We also analyzed the theoretical estimates of production rates for O($^{1}$S) and O($^{1}$D) atoms which depend on the water production rate \citep{Raghuram2013, Raghuram2014}. For comets 73P-C, 8P and 103P with Q$_{H_{2}O}$ $\sim$~10$^{28}$~s$^{-1}$, O($^{1}$S) comes both from CO$_{2}$ and H$_{2}$O while O($^{1}$D) is only produced by the photodissociation of H$_{2}$O. However, for C/2002 T7 (LINEAR) the O($^{1}$S) atoms are formed by the photodissociation of CO$_{2}$ close to the nucleus (<50~km) and, beyond, by both CO$_{2}$ and H$_{2}$O. Therefore, looking at the O($^{1}$S)/O($^{1}$D) values in Table~\ref{productionrates}, we should expect a higher G/R ratio below 50~km for this comet. Unfortunately, the atmospheric seeing does not permit us to resolve the coma at such small distances to the nucleus. Beyond $10^3$~km, the main contribution to both O($^{1}$S) and O($^{1}$D) is the photodissociation of H$_{2}$O molecules. As shown in Figs.~\ref{gr-linear}, \ref{gr-sw}, \ref{gr-tuttle} and \ref{gr-hart}, at such distances, the G/R ratio reaches a constant value of $\sim$0.05 which is very close to the O($^1$S)/O($^1$D) value for the pure water case (cf. Table~\ref{productionrates}). \\

\begin{table*}[htbp]
\begin{center}
\tiny
\tablefirsthead{
\hline
\hline
Comet & $r$ & $Q_{H_{2}O}$ ($10^{28}$ s$^{-1}$) & Date &  \hfill{Reference for $Q_{H_{2}O}$} \\
\hline
}
\tablelasttail{\hline}
\caption{Water production rates around the observing dates. Comets Hyakutake and Hale-Bopp included for comparison.}
\label{waterp}
\begin{supertabular}{p{4cm} >{\centering\arraybackslash}p{1cm} >{\centering\arraybackslash}p{2cm}>{\centering\arraybackslash}p{1.8cm}p{3.5cm}}
 C/2002 T7 (LINEAR) & 0.68 & 35.5 & 2004 May 5 &  \hfill{\cite{Disanti2006}} \\
  & 0.94 & 52.1 & 2004 May 27 &  \hfill{\cite{Combi2009}} \\
  73P-C/Schwassmann-Wachmann 3 & 0.95 & 1.70 & 2006 May 17 & \hfill{\cite{Schleicher2011}} \\
  8P/Tuttle & 1.04 & 1.4 & 2008 Jan 3 & \hfill{\cite{Barber2009}} \\
  103P/Hartley 2 & 1.06 & 1.15 & 2010 Nov 31& \hfill{\cite{Knight2013}} \\
  C/1996 B2 (Hyakutake) & 0.94 & 26 & 1996 Apr 1 & \hfill{\cite{Combi1998}} \\
  C/1995 A1 (Hale-Bopp) & 0.93 & 957 & 1997 Mar 26 & \hfill{\cite{DelloRusso2000}} \\ 
 \hline
\end{supertabular}
\end{center}
\end{table*}
\normalsize

\subsubsection{Estimation of CO$_{2}$ relative abundances}

We used the coupled-chemistry-emission model of \cite{Bhardwaj2012}
to estimate the CO$_2$ relative abundance (i.e. the CO$_{2}$/H$_{2}$O abundance ratio) in these comets by comparing the observed slit-centered G/R data points.
 The model accounts for the major production and loss processes of O($^{1}$S) and O($^{1}$D) atoms in the inner cometary coma.
 This model also includes the collisional quenching of O($^{1}$S) (4 $\times$ 10$^{-10}$ molecule$^{-1}$ cm$^{3}$ s$^{-1}$, \citealt{Stuhl1969}) and O($^{1}$D) (2.1 $\times$ 10$^{-10}$ molecule$^{-1}$ cm$^{3}$ s$^{-1}$, \citealt{Atkinson2004, Takahashi2005}) by H$_{2}$O in the inner coma. More details about the model calculations are given in \cite{Bhardwaj2012, Raghuram2013, Raghuram2014}. We take the atmospheric seeing effect into account, which is especially important very close to the nucleus. For this, we convolved the spatial profile provided by the model with a gaussian of FWHM equal to the seeing value. The seeing values are obtained from the Paranal seeing monitor.
 The H$_{2}$O production rates and the seeing values  used in the model
 as well as the  corresponding heliocentric distances 
  are given in Tables~\ref{observations} and \ref{waterp}. The CO relative abundances for comets C/2002 T7 and 73P-C are assumed equal to 5$\%$ in the model. However, the earlier works \citep{Bhardwaj2002, Raghuram2013, Raghuram2014} have shown that the role of CO in the determining G/R ratio is insignificant. The derived CO$_{2}$ relative 
 abundances obtained by fitting the observed G/R profiles with the model are 
 provided in the last column of Table~\ref{mod-input}.

\begin{figure}[h!]
\centerline{\includegraphics[width=\columnwidth]{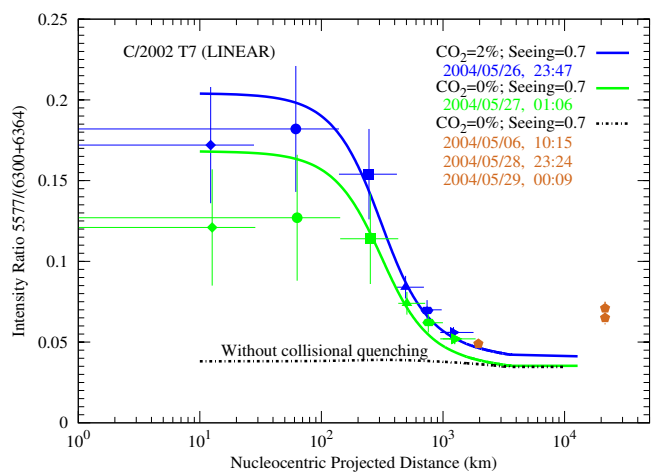}}
\caption{The G/R ratio for each sub-slit and offseted spectra for comet C/2002 T7 (LINEAR). The range of the nucleocentric distances covered by each point as well as the error on the G/R ratio are represented. The seeing is included in the x-errors and is indicated in km. It corresponds to the smallest size that can be resolved and explains the plateau found for the smallest spatial bin close to the nucleus. The calculated green to red-doublet emission intensity ratios as a function of the projected nucleocentric distance are plotted with solid curves. A H$_2$O production rate of 5.2 $\times$ 10$^{29}$ s$^{-1}$ and a 5\% CO relative abundance have been used with different seeing values and with different CO$_{2}$ relative abundances. The fits give the CO$_{2}$ relative abundance of the comet. Only non offset data points are considered for the computation of the G/R profile provided by the model. 
The black dash-dotted curve represents the calculated G/R ratio accounting for no collisional quenching with 0\% CO$_2$ relative abundance. The pentagon symbols are the data points for the offset observations.}
\label{gr-linear}
\end{figure}

\begin{figure}[h!]
\centerline{\includegraphics[width=\columnwidth]{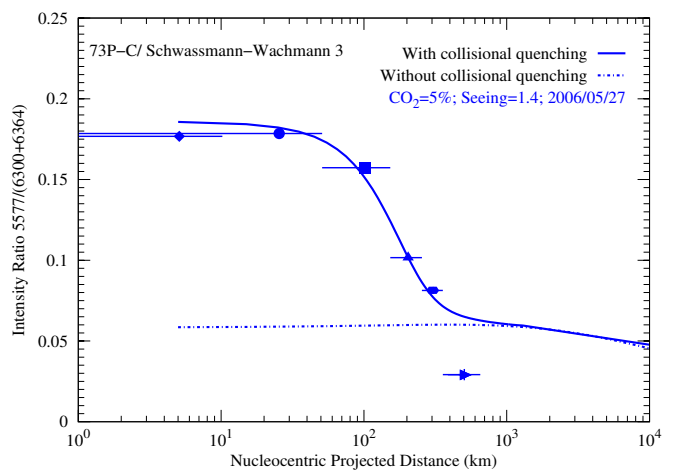}}
\caption{The G/R ratio for each sub-slit spectrum for comet 73P-C/Schwassmann-Wachmann 3. The range of nucleocentric distances covered by each point is indicated. The model calculated green to red-doublet emission intensity ratio as a function of projected distance is plotted with solid curve. A H$_2$O production rate of 1.7 $\times$ 10$^{28}$ s$^{-1}$ and 
 0.5\% CO relative abundance have been used. The fit gives the CO$_{2}$ relative abundance of the comet. The blue dash-dotted curve represents the calculated G/R ratio accounting for no collisional quenching with 0$\%$ CO$_{2}$ relative abundance.}
\label{gr-sw}
\end{figure}

\begin{figure}[h!]
\centerline{\includegraphics[width=\columnwidth]{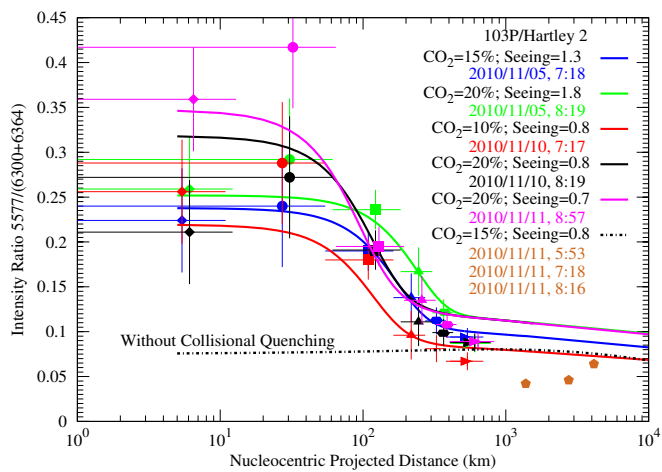}}
\caption{The G/R ratio for each sub-slit and offset spectra for comet 103P/Hartley 2. The range of nucleocentric distances covered by each point as well as the error on the G/R ratio are represented. The model calculated green to red-doublet emission intensity ratios as a function of projected distance are plotted. A H$_2$O production rate of 1.18 $\times$ 10$^{28}$ s$^{-1}$ and 0.5\% CO as relative abundance have been used with different seeing values and with different CO$_{2}$ relative abundances.
The fits give the CO$_{2}$ relative abundance of the comet.
The black dash-dotted curve represents the calculated G/R ratio accounting for no collisional quenching with 15\% CO$_2$ relative abundance. The pentagon symbols are the data points for the offset observations.}
\label{gr-hart}
\end{figure}

The comparison between the observed and the computed G/R  profiles of comet C/2002 T7 shown in Fig.~\ref{gr-linear} indicates that the  relative abundance of CO$_{2}$ in this comet is
less than 2\%. 
Similarly, the computed G/R profile for comet 73P-C with 5\% CO$_2$ relative abundance are presented in Fig.~\ref{gr-sw} together with the observations. 
The model calculations on comet 8P suggests that for 8P/Tuttle, the CO$_2$ relative abundance in this comet should be between 0 and 5\%. 
In comet 103P (shown in Fig.~\ref{gr-hart}), the required CO$_{2}$ relative abundances needed to reproduce the observations are about 15 to 20\% which is in agreement with the CO$_{2}$ measurements obtained by EPOXI observations \citep{AHearn2011}. 
This is the only comet with such a high CO$_{2}$ abundance in our sample. The observed data close to the nucleus are pretty well fitted with the model calculations on all these comets, which indicates that the O($^{1}$S) and O($^{1}$D) collisional quenching rates measured by \cite{Stuhl1969} and \cite{Takahashi2005} lie in a plausible range. \\

\begin{table*}[htbp]
\begin{center}
\tiny
\tablefirsthead{
\hline
\hline
Comet & $Q_{H_{2}O}^{a}$  & $r$ &  \multicolumn{2}{c}{Relative abundance (\%)} \\
 & ($10^{28}$ s$^{-1}$) & (au) & [CO] & \hfill{Model [CO$_{2}$]} \\
\hline
}
\tablelasttail{\hline}
\caption{The input parameters used in the model and the derived CO$_{2}$ abundances relative to H$_{2}$O.}
\label{mod-input} 
\begin{supertabular}{p{4cm} >{\centering\arraybackslash}p{1.5cm} >{\centering\arraybackslash}p{1cm} >{\centering\arraybackslash}p{1.5cm}p{1.5cm}}
 C/2002 T7 (LINEAR) & 52 & 0.94 & 5$^{b}$ & \hfill{0 - 2} \\
 73P-C/Schwassmann-Wachmann 3 & 1.7 & 0.95 & 5$^{b}$ & \hfill{5} \\
 8P/Tuttle & 1.4 & 1.03 & 0.5$^{c}$ & \hfill{0 - 5} \\
 103P/Hartley 2 & 1.15 & 1.06 & 0.15-0.45$^{d}$ & \hfill{15 - 20} \\
 \hline
\end{supertabular}
\end{center}
\tablefoot{$^{a}$ Taken from Table~\ref{waterp}\newline
$^{b}$ Assumed by Raghuram \& Bhardwaj model. \newline
$^{c}$ \cite{Bonev2008} \newline
$^{d}$ \cite{Weaver2011}}
\end{table*}
\normalsize

\subsubsection{103P/Hartley 2}

\begin{figure}[h!]
\centerline{\includegraphics[width=\columnwidth]{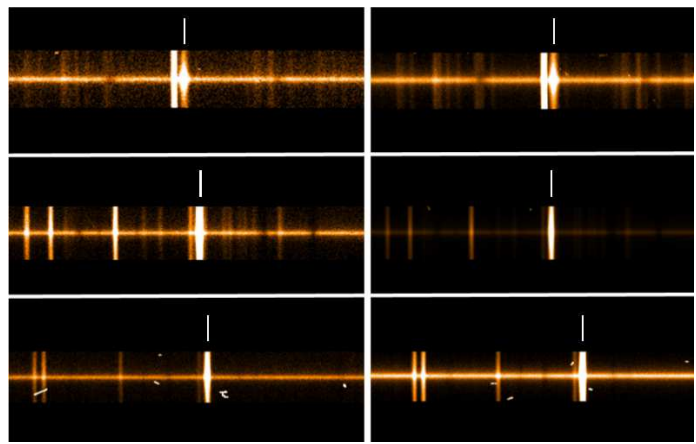}}
\caption{[OI] lines in 2D echelle spectra of 103P/Hartley 2, on 2010/11/10, 09:19 (right) and on 2010/11/11, 08:57 (left). From top to bottom, the 5577 $\AA$, 6300 $\AA$ and 6364 $\AA$ lines. The positions of the [OI] lines are indicated. No cosmic ray feature contaminates the [OI] lines.}
\label{cosmic}
\end{figure}

Two data points very close to the nucleus (observed on 2010/11/11, 8:57) appear relatively high in Fig.~\ref{gr-hart} while the other points further away from the nucleus have similar G/R values as the other spectra. First, we thought of a possible contamination by a cosmic ray but an analysis of the three [OI] lines in the 2D spectra shows no cosmic ray event (see Fig.~\ref{cosmic}). Fig.~\ref{cosmic} also confirms that there was no problem during the acquisition of this spectrum by comparing it to the one taken the day before. This suggests that the CO$_{2}$/H$_{2}$O abundance could sometimes strongly vary close to the nucleus. These variations could be due to icy particles that evaporate very fast in the first ten kilometers and/or to the rotation of the nucleus, as it has been shown that the CO$_{2}$ is coming from a localized region (the small lobe) of the nucleus \citep{AHearn2011}.

\subsection{Line widths}

\begin{figure}[h!]
\centerline{\includegraphics[width=\columnwidth]{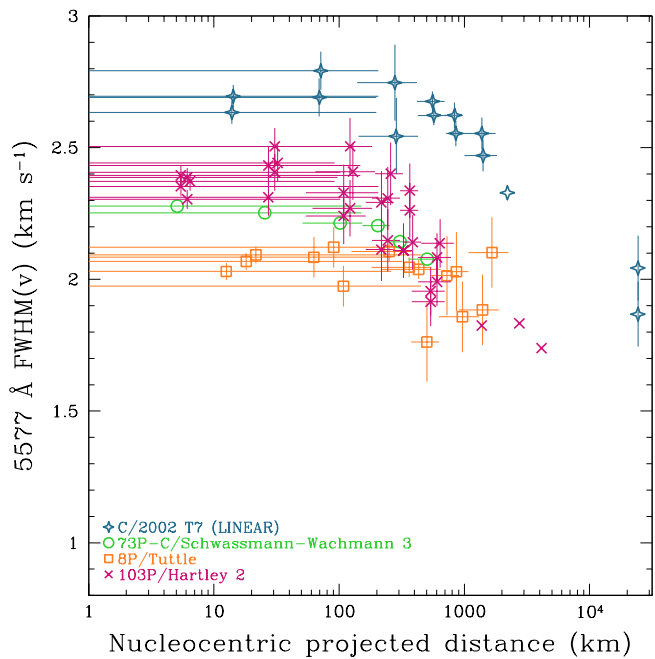}}
\caption{The 5577.339 $\AA$ line width against the distance to the nucleus for all comets. The errors on the distance and the widths are represented. A decrease of the width with the nucleocentric distance is observed for all comets. Within about 300~km, there is a plateau because of the convolution by the seeing.}
\label{fwhmgreen}
\end{figure}

\begin{figure}[h!]
\centerline{\includegraphics[width=\columnwidth]{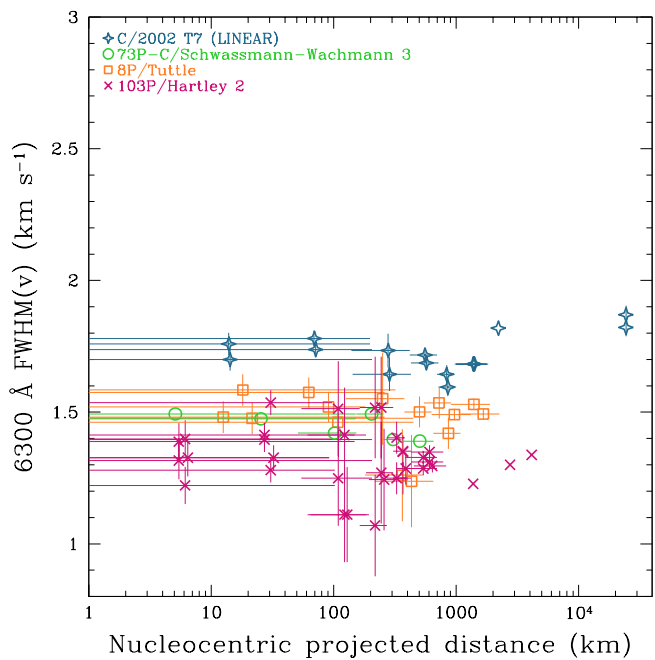}}
\caption{Same figure as Fig.~\ref{fwhmgreen} but for the 6300.304 $\AA$ line.}
\label{fwhmred1}
\end{figure}

\begin{figure}[h!]
\centerline{\includegraphics[width=\columnwidth]{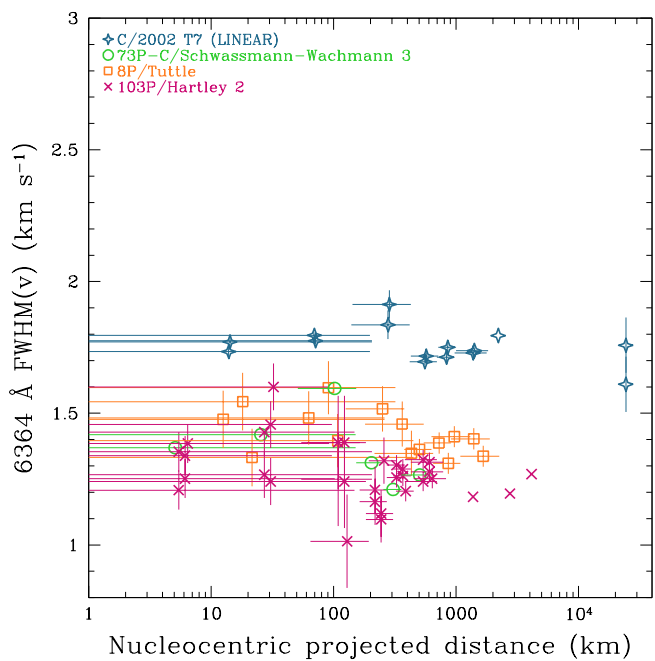}}
\caption{Same figure as Fig.~\ref{fwhmgreen} but for the 6363.776 $\AA$ line.}
\label{fwhmred2}
\end{figure}

Using relation (9) given in \cite{Decock2013},

\begin{eqnarray}
\rm{FWHM(v)} = \frac{FWHM_{intrinsic}~(\rm{\lambda})~~\rm{c}}{2~\lambda_{[OI]}~\sqrt{\ln~2}}~,
\label{width}
\end{eqnarray}
where the intrinsic line widths (FWHM($\lambda$)), corrected for the instrumental broadening, are evaluated from the measured FWHM$_{\rm{observed}}$ ($\AA$). The results are listed in Table~\ref{tableaulong} and the errors on the velocity widths are provided in Table~\ref{average}. Figures~\ref{fwhmgreen}, \ref{fwhmred1} and \ref{fwhmred2} plot the intrinsic width of the three forbidden oxygen lines as a function of the nucleocentric distance. In our observations, we clearly notice that the green line is wider than the red lines ($\sim$2.5~km~s$^{-1}$ versus $\sim$1.5~km~s$^{-1}$) even after the subtraction of the C$_{2}$ contamination. This was first noticed by \cite{Cochran2008} and confirmed by \cite{Decock2013} for a large sample of comets. It should be emphasized that 73P-C presents, like all comets, a wider green line while this comet is very poor in C$_{2}$. Therefore, the larger width of the $5577~\r{A}$ line cannot be explained by the C$_{2}$ blends alone. A blend with another species cannot be excluded but this is unlikely. The green line width decreases when the distance to the nucleus increases (Fig.~\ref{fwhmgreen}). This trend is more obvious for comets C/2002 T7 and 103P for which we have spectra taken at large distances from the nucleus. The green line width at large distance is also getting similar to the width of the red lines. 
Extrapolating the linear regression for the green line width of 103P and C/2002 T7 with the distance, we expect that the green line width will be equal to the red line width at $\sim$5000~km for 103P
and $\sim$35000~km for C/2002 T7. More observations at larger distances from the nucleus are needed to confirm it. We cannot explain this behavior by the collisional quenching because this is not seen for the red lines. Indeed, as already mentioned by \cite{Raghuram2014}, the O($^{1}$D) atoms are more quenched than the O($^{1}$S) atoms due to their longer lifetime. Therefore, the collisional quenching should make the red line wider than the green line which is not observed.
The larger width of the green line close to the nucleus could be explained by the fact that the O($^{1}$S) atoms are mainly produced by CO$_{2}$ very close to the nucleus. Indeed, the solar flux responsible for the production of O($^{1}$S) from CO$_{2}$ comes from the 955-1165 $\AA$ region \citep{Bhardwaj2012} which is more energetic than the Lyman-$\alpha$ photons, supposedly considered as the main source of oxygen atom production from H$_{2}$O dissociation. These energetic photons can penetrate deeper in the thick coma than the Lyman-$\alpha$ photons. Since the line width is related to the value of the excess energy remaining after the photodissociation, an excitation by higher energetic photons produces larger widths, as we notice for the green line close to the nucleus. The excess energy of different parent molecules has been evaluated by \cite{Raghuram2014}. They found that the excess energy from CO$_{2}$ is higher than that from H$_{2}$O. When the O($^{1}$S) atoms are observed far away from the nucleus, they are coming from both CO$_{2}$ and H$_{2}$O but with Lyman-$\alpha$ photons as the major excitation source, and then with a larger contribution from the water molecules, which results in a decrease of the green line widths. \\
\cite{Bisikalo2014} investigated about the larger width of the green line and provided a different explanation. They developed a Monte Carlo model to study the [OI] lines widths and suggested that the observed line profile also depends on the thermalization due to elastic collisions. Nevertheless, the thermalization process leads to a broadening of the red lines and not of the green line which is not in agreement with our observations.\\
In Figs.~\ref{fwhmgreen} to \ref{fwhmred2}, we notice that the widths of the three forbidden oxygen lines of C/2002 T7 (LINEAR) are always larger than the ones of the other comets. We explain this by the higher water production rate of this comet. Indeed \cite{Tseng2007} have shown that the expansion velocity of the molecules increases when the water production rate increases.

\section{Conclusions}

Four comets have  been observed at small geocentric distances with the UVES spectrograph (VLT). 17 high-resolution spectra have been collected when the comets were very close to the Earth (< 0.65~au) allowing the study of the [OI] lines as close as 100~km from the nucleus. Some of them were also recorded with an offset from the nucleus corresponding to a distance of 10~000~km. The 11 centered spectra were spatially extracted and binned per pixel zones in order to get 1D spectra at various distances from the nucleus. Finally, we studied the forbidden oxygen lines in 63 sub-spectra and 6 offset spectra corresponding to different nucleocentric distances. The G/R ratio and the velocity widths have been computed and analyzed in order to better understand the production of oxygen atoms in the coma. The results can be summarized as follows :
\begin{enumerate}
\item In this analysis, C$_{2}$ blends have been identified and removed from the oxygen green line. This subtraction, made for the first time, is important because this contamination modifies the intensity and the FWHM  of the 5577.339~$\AA$ line for comets rich in C$_{2}$ and more specifically for the analysis made far away from the nucleus\footnote{Note that the C$_{2}$ correction has no influence on our previous results \citep{Decock2013}.}. However, the C$_{2}$ blends do not explain the larger width of the green line compared to the red ones.
\item Thanks to the high spatial resolution of the data, we found that the G/R ratio as a function of the nucleocentric distance displays the same profile for all the comets: a rapid increase below 1000~km up to typically 0.2, and a constant value of about 0.05 at larger distances. The G/R value of about 0.1 found in previous studies is thus an average of the values obtained both close to the nucleus and at larger distances. This particular profile is mainly explained by the quenching  which plays an important role in the destruction of O($^{1}$D) atoms in the inner coma (<1000~km). The parent molecules forming oxygen atoms can also contribute to this trend. Indeed, O($^{1}$D) atoms are only formed by H$_{2}$O while O($^{1}$S) are produced by both CO$_{2}$ and H$_{2}$O. With increasing distance from the nucleus, the relative H$_{2}$O contribution becomes larger and thus the intensity of the green line decreases.
In the extended coma, oxygen atoms are only formed by the photodissociation of water molecules. The G/R value of 0.05 is in good agreement with the pure water case (Table~\ref{productionrates}).
\item We fitted our observational [OI] emission lines data with coupled-chemistry-emission model calculations. The comparison between observations and model calculations suggest that the collisional quenching of O($^{1}$S) and O($^{1}$D) by H$_{2}$O molecules is significant in the inner coma and cannot be neglected while determining the G/R ratio. Using the measured G/R values close to the nucleus, an estimate of the relative abundance of CO$_{2}$ could also be derived for all the comets of the sample. 
\item The green line has a much larger width, typically 2.5 km~s$^{-1}$ compared to 1.5~km~s$^{-1}$ found for the red lines. The higher width of the green line is likely due to the involvement of different parent molecules in producing the O($^{1}$S) and O($^{1}$D) atoms (as already suggested by \cite{Bhardwaj2012, Decock2013}): the green oxygen lines are produced by oxygen atoms coming from the photodissociation of both the CO$_{2}$ and H$_{2}$O molecules while the red ones are produced mainly by H$_{2}$O molecules. While the O($^{1}$S) atoms are mostly provided by CO$_{2}$, the contribution of water molecules increases with the nucleocentic distance. This explains the decreasing width of the 5577.339 $\AA$ line at distances larger than 500~km.
\item Measuring the green and red-doublet emission intensities and their line widths is a new way to constrain the CO$_{2}$ relative abundances in comets.
\end{enumerate}

\begin{acknowledgements}
A.D acknowledges the support of the Belgian National Science Foundation F.R.I.A., Fonds pour la formation \`a la Recherche dans l'Industrie et l'Agriculture. E.J. is Research Associate FNRS, D.H. is Senior Research Associate at the FNRS and J.M. is Research Director of the FNRS. S.R. was supported by ISRO Research associateship during a part of this work. The work of A. B. was supported by ISRO. C. Arpigny is thanked for helpful discussions and important constructive comments.
\end{acknowledgements}

\bibliographystyle{aa}
\bibliography{article_nucleus}
\end{document}